# Experimental Autoignition of $C_4$-$C_6$ Saturated and Unsaturated Methyl and Ethyl Esters

H. Bennadji*, J. Biet, L. Coniglio-Jaubert, F. Billaud, P.A. Glaude, F. Battin-Leclerc

Département de Chimie-Physique des Réactions, UMR n°7630 CNRS, INPL-ENSIC
1, rue Grandville - BP 451 - 54001 Nancy Cedex – France

**Abstract**
Autoignition delay times, $\tau$, of methyl crotonate, methyl acrylate, ethyl butanoate, ethyl crotonate, and ethyl acrylate were studied in shock tube experiments. A series of mixtures diluted with argon, of varying fuel/oxygen equivalence ratios ($\Phi$=0.25, 0.4, 1.0, and 2.0), were measured behind reflected shock waves over the temperature range of 1280-1930 K, pressure range of 7-9.65 atm, during which the logarithm of $\tau$ varies linearly as a function of the inverse temperature for all equivalence ratios. The ignition delay time decreases as temperature rises. The dependence of $\tau$ on temperature, and reactant concentrations is given in an empirical correlation. The results provide a database for the validation of small saturated and unsaturated esters kinetic mechanisms at elevated temperatures and pressure combustion.

## 1. Introduction

Biodiesel is an alternative fuel which can be used directly to a diesel engine without modifying the engine system. Basically, biodiesel can be attained by transesterification of oils and fat that come from oleaginous plants, used vegetal oils, and animal fat [1]. It has been the focus of a considerable amount of recent research because it is renewable and reduces the emission of some pollutants. Its production and use has increased significantly in many countries around the world, including the United States, Austria, France, Germany, Italy, and it is in nascent status in many others. It has the major advantages of having high biodegradability, excellent lubricity and no sulfur content [2]. It can be stored just like mineral diesel and hence does not require separate infrastructure. The use of biodiesel in conventional diesel engines results in substantial reduction in emission of unburned hydrocarbons, carbon monoxide and particulate [3]. Some technical problems facing biodiesel include the reduction of NOx exhaust emissions and the improvement of cold flow properties among such as oxidative stability and economics. Nevertheless, increasing cetane number will cause significant reductions in the NOx emissions due to shorter ignition delay times and the resulting lower average combustion temperatures [4].

Biodiesel contains from 10 wt% to 12 wt% oxygen, which lowers energy density and the particulate emission [2]. Furthermore, saturated and unsaturated methyl or ethyl esters are the mainly compounds of biodiesel fuels, containing carbon chains of twelve or more atoms in length [5]. Because of this chemical structure, both their experimental and simulation studies are difficult. Circumvention of this problem is by selecting of a surrogate molecule containing the same chemical functional group as real biodiesel fuel [6]. Few experimental and kinetic data are available in the literature about the oxidation of oxygenated compounds such as esters. As biodiesel surrogate, methyl butanoate (MB) has been the target of many published studies [6-13]. The first detailed mechanism for the combustion of MB has been developed by Fisher et al. [6], including 264 species and 1219 reactions. This mechanism has been used by Marchese et al. [7] to simulate their results obtained in a flow reactor at 12.5 atm, over the temperature range 500-900 K and at different equivalence ratios of 0.35-1.5. Later, Huynh and Violi [8] developed an improved MB pyrolysis model by using ab initio technique. Other recent investigations have also attempted to improve MB kinetic mechanism as approach other model biodiesel molecules. Gail et al. [9] modified the Fisher's mechanism and investigated the MB combustion in a jet stirred reactor, an opposed flow diffusion flame, and a Princeton variable pressure flow reactor. Their new numerical model consists of 295 species and 1498 reactions. Metcalfe et al. [10] have published comparative investigation on the oxidation of MB and ethyl propanoate (EP) in shock tube at 1 and 4 atm, over the temperature range 1100-1670 K. It was found that EP is faster to ignite than MB. Following this work, measurements of ignition delay for MB have been completed by Dooley et al. [11]. The autoignition data were obtained from rapid compression machine (RCM) over the temperature range 640-949 K at compressed gas pressures of 10, 20, and 40 atm at varying equivalence ratios of 1.0, 0.5, and 0.33. Farooq et al. [12] studied the pyrolysis of three methyl esters: methyl acetate, methyl propionate, and MB behind reflected shock tube by measuring $CO_2$ time-histories over temperature range from 1260 to 1653 and at pressure from 1.4 to 1.7 atm. It found that the $CO_2$ yields were not strongly dependent on the alkyl chain length. The combination of the Fisher et al. model [6] and the theoretical work of Huynh and Violi [8] accurately recover the experimental measured $CO_2$ yields. However, very few detailed kinetic data have been reported for unsaturated esters. Sarathy et al. [13] studied methyl crotonate (MC) in a comparative experimental study of its oxidation and a saturated methyl ester, MB. The experiments were carried out in an opposed flow diffusion flames for a mixture containing 4.7% fuel, 42% $O_2$, 58% $N_2$ and in jet stirred reactor at a residence time of 70 ms, over the temperature range 850-1350 K for a mixture of 0.075% fuel under stoichiometric conditions at 1 atm pressure. Mole fraction profiles of major intermediates and final products were measured together with that of the reactants. They concluded that both fuels have similar reactivity and proposed that unsaturated esters would have more tendencies to soot formation than saturated esters. Furthermore, recently, Gaïl et al. [14] provided new experimental results for the oxidation of MC which were performed in a jet stirred reactor at 1 atm pressure, over the temperature 850-1400 K, under two equivalence ratios $\Phi$ =0.375, 0.75 with a residence time of 70 ms. In addition, a new detailed kinetic mechanism for MC has been proposed by analogy with reactions in the MB mechanism [9]. Overall, their kinetic model reproduced the experimental data fairly well.

However, no studies about the ignition delay times of the unsaturated small esters were performed behind reflected shock tube, as well as the effect of the double bond on the autoignition are even scarce. The aims of this study are to provide more information on the combustion characteristics of esters as model of biodiesel, as well as to compare the reactivity between methyl and ethyl esters, and the effect of the double bond on the autoignition of small esters.

## 2. Experimental

The ignition delay time, $\tau$ is an important chemical kinetic characteristic of combustion compounds. In this study, experiments were performed in reflected shock tube setup which was reported in detail elsewhere ([15], [16], [17]) and only a brief description is given here. Autoignition delay times have been measured in a stainless steel shock tube; the reaction and the driver parts were respectively 400.6 and 90 cm in length and 7.8 and 12.82 cm in diameter and were separated by two terphane diaphragms. These diaphragms were ruptured by suddenly decreasing the pressure in the space separating them, that allowed us to keep the same pressure (5.3 bar) in the high pressure part for all experiments. The driver gas was helium. The incident and reflected shock velocities were measured by four piezo-electric pressure transducers located along the reaction section. By using the same notations as [15, 16, and 17], the pressure $P_5$ and temperature $T_5$ of the test gas behind the reflected shock wave were derived from the values of the initial pressure in the low pressure section (ranging from 10 to 40 kPa) and of the incident shock velocity by using ideal one-dimensional shock equations. The error on the temperature was about 20 K.

The onset of ignition was detected by OH* radical emission at 306 nm through a quartz window with a photomultiplier fitted with a monochromator at the end of the reaction part (the window was located at 2 mm of the end-plate of the tube). The quartz window was located at the same place along the axis of the tube as the last pressure transducer. Fresh reaction mixtures were prepared daily in a 20 L stainless steel tank using standard manometric methods. Before each introduction of the reaction mixture, the reaction section was flushed with pure argon and evacuated, to ensure the residual gas to be mainly argon. Research grade Argon, Helium, and Oxygen specified to be 99.995% pure, were supplied by Messer and were used without further purification. Methyl and ethyl esters (>99% pure) were supplied by Sigma-Aldrich Co. As shown in Fig. 1, $\tau$ was defined as the time interval between the pressure rise measured by the last pressure transducer due to the arrival of the reflected shock wave and the rise of the optical signal by the photomultiplier up to 50% of its maximum value. The mixtures and the conditions for this investigation were selected to understand and predict the effect of the concentration, the equivalence ratios on the ignition delay times for each fuel studied. In addition, to compare the ignition properties of the methyl and ethyl esters taking in account the effect of the double bond. The thermodynamic data for each ester were generated in the form of NASA polynomials using THERGAS software [18], based on Benson's group additivity method [19].

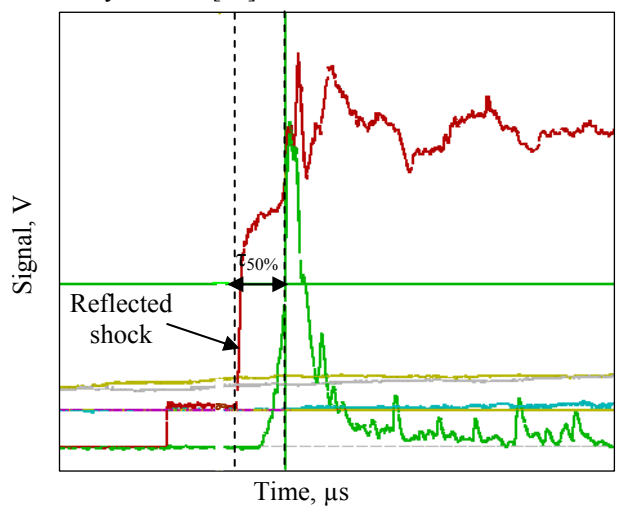

Fig.1. Determination of the ignition delay time using OH* emission diagnostic and pressure traces from an experiment at $\Phi$=0.25 and mixture of 1% MC, 6% $O_2$, and 93% Ar.

Table 1
Methyl and ethyl esters included in this study

| Species | Formula | Structure |
|---|---|---|
| Methyl Crotonate (MC) | $C_5H_8O_2$ | $CH_3CH=CH(C=O)OCH_3$ |
| Methyl Acrylate (MAC) | $C_4H_6O_2$ | $CH_2=CH(C=O)OCH_3$ |
| Ethyl Butanoate (EB) | $C_6H_{12}O_2$ | $CH_3CH_2CH_2(C=O)OCH_2CH_3$ |
| Ethyl Crotonate (EC) | $C_6H_{10}O_2$ | $CH_3CH=CH(C=O)OCH_2CH_3$ |
| Ethyl Acrylate (EAC) | $C_5H_8O_2$ | $CH_2=CH(C=O)OCH_2CH_3$ |

Table 2
Mixture compositions, shock conditions, and ignition delay times

| Fuel | $\Phi$ | Test gas composition | | | $P_5$ (atm) | $T_5$ (K) | $\tau$ (μs) |
|---|---|---|---|---|---|---|---|
| | | X (%) | $XO_2$ (%) | $X_{Ar}$ (%) | | | |
| MC | 0.25 | 0.5 | 12 | 87.5 | 7.2-8.35 | 1346-1510 | 7.33-90.8 |
| | 1 | 0.5 | 3 | 96.5 | 7.6-9 | 1453-1593 | 18.32-91.12 |
| | 1 | 1 | 6 | 93 | 7.85-8.82 | 1305-1582 | 17.34-348 |
| | 2 | 0.5 | 1.5 | 98 | 7.14-8.48 | 1476-1800 | 10.96-178.8 |
| MAC | 0.25 | 0.5 | 9 | 90.5 | 7.6-8.74 | 1388-1521 | 4.05-46.3 |
| | 1 | 0.5 | 2.25 | 97.25 | 7-8.27 | 1434-1624 | 7.14-75.4 |
| | 1 | 1 | 4.5 | 94.5 | 7.26-9.62 | 1341-1510 | 10.25-205 |
| | 2 | 0.5 | 1.125 | 98.38 | 7.31-8.58 | 1485-1765 | 16.9-95 |
| EB | 0.25 | 0.5 | 16 | 83.5 | 7.75-8.8 | 1280-1454 | 4.2-188.6 |
| | 1 | 0.42 | 3.33 | 96.25 | 7.65-9.22 | 1307-1740 | 5.5-1485 |
| | 1 | 0.5 | 4 | 95.5 | 7.75-8.5 | 1385-1635 | 2.3-112 |
| | 1 | 1 | 8 | 91 | 7.8-9.15 | 1296-1474 | 10.95-321.5 |
| | 2 | 0.5 | 2 | 97.5 | 7-8.55 | 1532-1922 | 3.5-88.5 |
| EC | 0.25 | 0.5 | 16 | 83.5 | 7.5-8.95 | 1357-1627 | 6.4-62.9 |
| | 1 | 0.42 | 3.13 | 96.45 | 6.7-8.6 | 1411-1817 | 4.8-176.5 |
| | 1 | 0.5 | 3.75 | 95.75 | 7.5-8.5 | 1350-1653 | 22-242 |
| | 1 | 1 | 7.5 | 91.5 | 7.8-9.33 | 1284-1524 | 4.7-138 |
| | 2 | 0.5 | 1.875 | 97.63 | 7.24-8.4 | 1402-1885 | 12.23-240 |
| EAC | 0.25 | 0.5 | 12 | 87.5 | 7.75-9 | 1347-1517 | 4.2-19.6 |
| | 1 | 0.4 | 2.4 | 97.2 | 7.6-8.97 | 1404-1835 | 4.37-86.2 |
| | 1 | 0.5 | 3 | 96.5 | 7.74-9.2 | 1389-1688 | 7.45-45.84 |
| | 1 | 1 | 6 | 93 | 7.75-9.33 | 1328-1575 | 5.22-96.7 |
| | 2 | 0.5 | 1.5 | 98 | 7.42-8.46 | 1509-1833 | 4-55.4 |

*Note*: All mixture composition data are provided on a mole basis. The equivalence ratio ($\Phi$) is calculated based on the C-O molar ratios of the experimental and stoichiometric conditions. $P_5$ and $T_5$ are the pressure and temperature immediately behind the reflected shock wave, $\tau$ is the ignition delay time.

## 3. Results and Discussion

### 3.1. Ignition times

In this work, ignition times $\tau$ were measured from shock tube experiments using $OH^*$ radical emission at 306 nm over a wide range of temperatures, pressures, fuel concentrations, and equivalence ratios as showed in Table 2. Figs. 2 to 6 display the evolution of $\tau$ as a function of the temperature inverse for each ester. As can be seen, increasing the temperature and the oxygen molar fraction, with a constant ester molar fraction (0.5%), involve shorter $\tau$. This is understandable since the increase of oxygen concentration can lead to a significant increase in the concentration of O and OH radicals under lean conditions. The experimental data also show that the $\tau$ decreases as the esters concentration increases from 0.4% ($\Phi$ =1.0) to 1.0% ($\Phi$ =1.0). The same general trends have been found in many hydrocarbon and esters ignition delay time measurements [10].

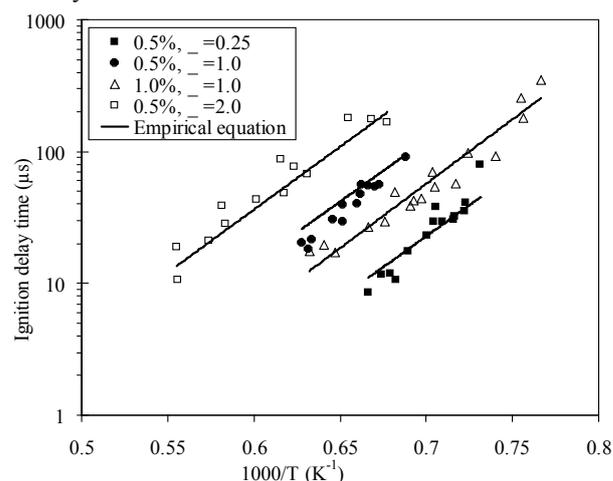

Fig. 2. Ignition delay times of MC versus reciprocal temperature behind reflected shock tube. Symbols are experimental data. Solid line represents ignition delay times calculated from correlation.

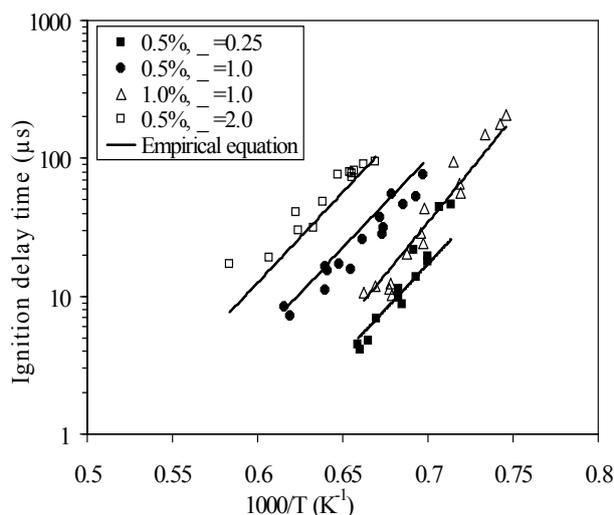

Fig. 3. Ignition delay times of MAC versus reciprocal temperature behind reflected shock tube. Symbols are experimental data. Solid line represents ignition delay times calculated from correlation.

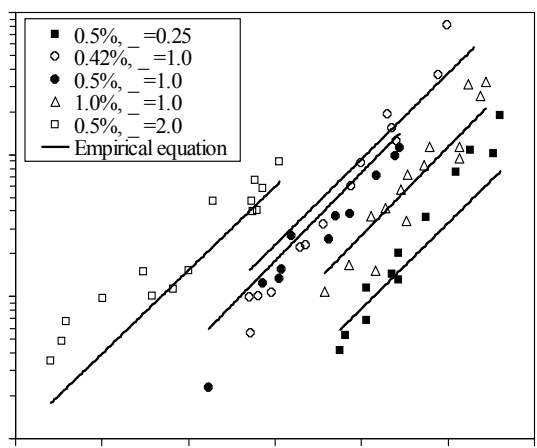

Fig. 4. Ignition delay times of EB versus reciprocal temperature behind reflected shock tube. Symbols are experimental data. Solid line represents ignition delay times calculated from correlation.

### 3.2. Ignition times correlation equation

From a regression analysis of all the present data, ignition times were found to scale with temperature, oxygen, and fuel concentrations, this power law dependence has been applied to present the variation of ignition time as an empirical Arrhenius type correlation: $\tau = A\exp(-E/RT)[\text{ester}]^a[O_2]^b$ (see table 3). The temperature and pressure ranges of the different experiments, and measured $\tau$ are shown in Table 2. Similar values of the activation energy for the MC and EC as for MAC and EB were found, ranging from 51.25 to 69.48 K cal/mol. However, that of EAC has different order of magnitude compared to the others. Besides, equations derived for this study show that the MC and EAC power dependence coefficients (a) are positive while those of MAC, EB, and EC are negative and the $O_2$ power dependence coefficients (b) are negative. A plot such as this can be seen in Figs. 2 to 6

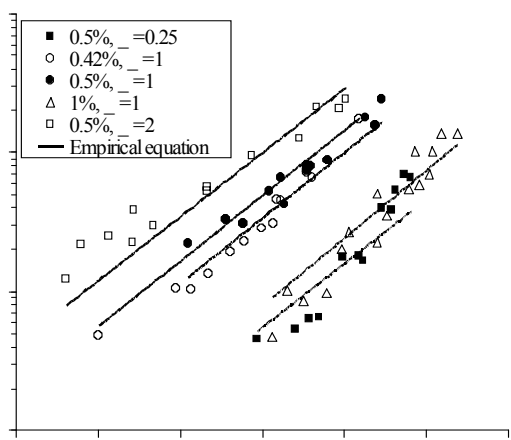

Fig. 5. Ignition delay times of EC versus reciprocal temperature behind reflected shock tube. Symbols are experimental data. Solid line represents ignition delay times calculated from correlation.

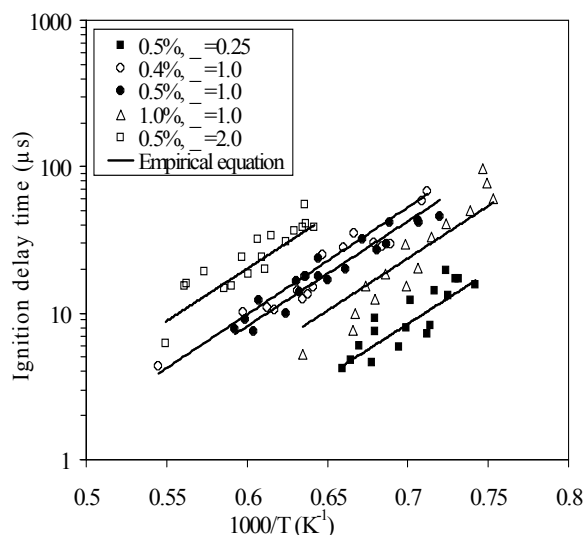

Fig. 6. Ignition delay times of EAC versus reciprocal temperature behind reflected shock tube. Symbols are experimental data. Solid line represents ignition delay times calculated from correlation.

which present a statistical goodness of fit, or $R^2$ value, 0.99.

Table 3
Summary of correlations found by fitting experimental data using the form $\tau = A\exp(E/RT)[\text{ester}]^a[O_2]^b$

| Ester | Parameters | | | |
|---|---|---|---|---|
| | **A** | **E** | **a** | **b** |
| MC | 3.19E-21 | 51.25 | 0.01 | -1.37 |
| MAC | 9.22E-28 | 69.48 | -0.47 | -1.36 |
| EB | 2.26E-26 | 65.80 | -0.04 | -1.76 |
| EC | 1.96E-28 | 54.89 | -0.79 | -1.5 |
| EAC | 4.98E-17 | 37.98 | 0.26 | -1.24 |

*Note*: Concentrations ([ ]) are in mol/cm$^3$, $\tau$ is in seconds, the activation energy (E) is in Kcal/mol, and R is the universal gas constant.

## 3.3. Comparison between esters

The comparisons between esters are plotted in Figs. 7 to 9. Ignition delay times for a saturated ethyl ester (EB) are compared with an unsaturated ethyl ester (EC) at stoichiometric mixture of 0.5% fuel (see Fig. 7). It is seen that EB ignites faster than EC. This difference in reactivity can be attributed to the effect of the double bond β-position of the ester function which gives more stability to ignite for the conjugated and unsaturated esters than those saturated ones. However, Saratay et al. [13] have studied oxidation of MC and MB in comparative investigation in an opposed flow diffusion flames and in jet stirred reactor. Their experiments results indicate that both fuels have similar reactivity. Furthermore, as shown in Fig. 8, the unsaturated fuels, MC and EC are compared for $\Phi =1$ and at the same molar fraction of carbon atom (2.5%). It was found that both fuels have similar reactivity. By contrast, the reactivity comparison between MAC and EAC (see Fig. 9) under shock tube conditions of $\Phi =1$ and at the same molar fraction of carbon atom (2%), shows that the behavior of each fuel varies slightly depending on the temperature domain: at higher temperatures, greater than 1500 K, MAC is more reactive than EAC. However, at lower temperatures, below 1500 K, EAC ignites faster than MAC. Nevertheless, Metcalfe et al. [10] have published comparative investigation on the oxidation of two saturated esters, MB and EP in shock tube at 1 and 4 atm, and over the temperature range 1100-1670 K. It was found that EP is faster to ignite than MB. This phenomenon is due to the six-centered unimolecular decomposition of EP into ethylene and propanoic acid. Consequently, the reactivity of these two products is responsible for the faster ignition of EP. Recently, Walton et al. [20] performed ignition of MB and EP using a rapid compression facility over the temperature range 935-1117 K, at pressure of 4.7-19.6 atm, and over equivalence ratios of $\Phi =0.3$-0.4. They also found that that EP ignites more rapidly than MB. This reactivity has explained by faster unimolecular decomposition of EP leading to the formation of ethylene and propanoic acid as showing by the works of Schwartz et al. [21] and Metcalfe et al. [10].

## 3.4. Comparison with chemical kinetic model

The detailed chemical kinetic mechanisms oxidation for the esters investigated is currently under development by using EXGAS [23], a computer package developed to perform the automatic generation of detailed kinetics models for the gas-phase oxidation. A small portion of this study has been devoted to the comparison of the experimental data against existing chemical kinetics models. Therefore, the comparison was only made to the chemical kinetic model of the MC ester which consists of 1516 elementary reactions and 301 species proposed early by Gaïl et al. [14]. Their model has been validated at atmospheric pressure and temperature range below 1300 K under jet stirred reactor and counterflow flame. Simulations have been achieved using the SENKIN software package of CHEMKIN II [24]. In general, as we can see in Fig.10,

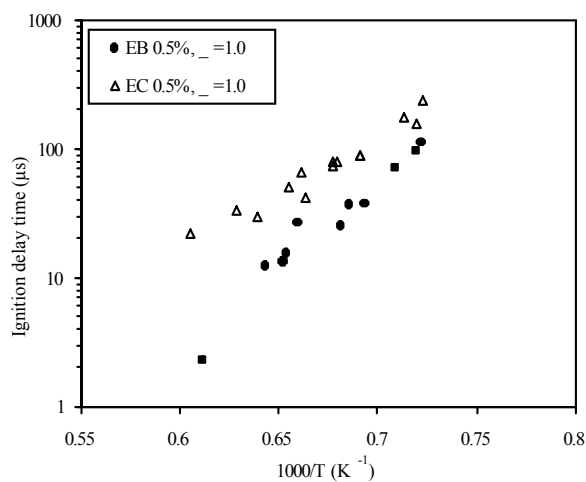

Fig. 7. Effect of unsaturation on ignition delay times: Comparison between EB and EC ignition delay times for stoichiometric mixture ($\Phi =1$) and 0.5% of fuels.

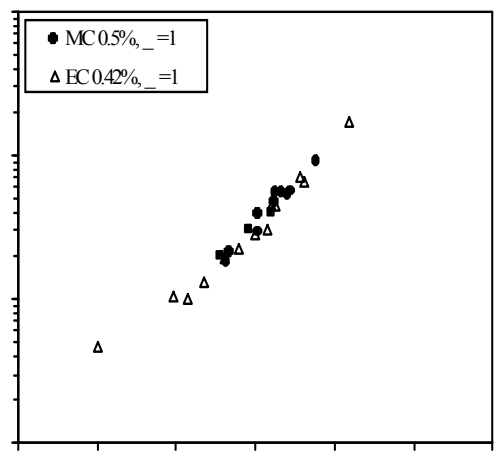

Fig. 8. Comparison between MC and EC ignition delay times at equal molar fraction of carbon atom (2.5%).

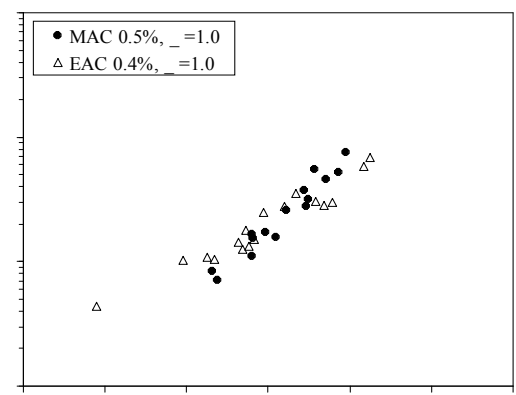

Fig. 9. Comparison between MAC and EAC ignition delay times at equal molar fraction of carbon atom (2%).

the mechanism shows a good agreement with measured ignition delay times of stoichiometric ($\Phi =1$) and rich

($\Phi$ =2) mixtures at 0.5 % fuel. However, it overpredicts the experimental data of lean mixture ($\Phi$ =0.25) by a factor of 2-3.

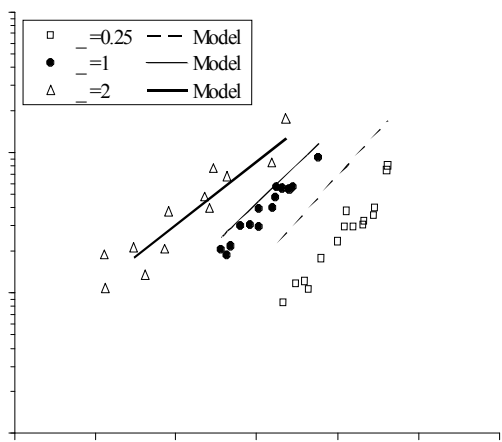

Fig. 10. Comparion of modeling and measured ignition delay times of MC, for mixtures at $X_{fuel}$=0.5% and different equivalence ratio. Symbols are experimental data. Lines are modeling data.

## 4. Conclusions

An ignition time database for small esters ignition has been generated at high temperatures and high pressure from shock tube experiments using optical diagnostic from excited OH* emission at 306 nm. For each fuel a global correlation for ignition delay time applicable over a wide experimental range has been proposed. The detailed chemical kinetic mechanisms oxidation for the esters investigated here is currently under development by using EXGAS. Furthermore, methyl crotonate oxidation under shock tube conditions was modeled using a detailed kinetic reaction mechanism (301 species and 1516 elementary reactions) proposed recently. Overall, acceptable agreements was obtained between experimental data and the computations for stoichiometric and rich mixtures at 0.5% of ester, the lean mixture ($\Phi$ =0.25) showed some disagreement. This new experimental data presented in this study provide critical kinetic targets to evaluate model performance, and should prove useful for researchers engaged in kinetic model development of biodiesel oxidation.